\def\be{\begin{equation}}
\def\ee{\end{equation}}
\def\ba{\begin{array}}
\def\ea{\end{array}}

\documentclass[preprint,showpacs,amsmath]{revtex4}
\usepackage{amsfonts}
\usepackage{amssymb}
\def\qed{\leavevmode\unskip\penalty9999 \hbox{}\nobreak\hfill
     \quad\hbox{\leavevmode  \hbox to.77778em{%
               \hfil\vrule   \vbox to.675em%
               {\hrule width.6em\vfil\hrule}\vrule\hfil}}
     \par\vskip3pt}

\newtheorem{theorem}{Theorem}

\input amssym.def
\begin{document}
\title{\large\bf A note on quantum Bell nonlocality and quantum entanglement for high dimensional quantum systems}
\author{Tinggui Zhang$^{1,\dag}$, Ya Xi$^2$ and Shao-Ming Fei$^{3,4}$\\[10pt]
\footnotesize
\small$1$ School of Mathematics and Statistics, Hainan Normal University,\\
\small Haikou 571158, P. R. China\\
\small$2$  Department of Mathematics, South China University of Technology, GuangZhou 510640, China\\
\small$3$ Max-Planck-Institute for Mathematics in the Sciences, Leipzig 04103, Germany\\
\small$4$ School of Mathematical Sciences, Capital Normal University, Beijing 100048, China  \\
\small $^\dag$ Correspondence to tinggui333@163.com}
\bigskip

\begin{abstract}
{\bf Abstract}  We study the Bell nonlocality of high dimensional
quantum systems based on quantum entanglement. A quantitative
relationship between the maximal expectation value $B$ of Bell
operators and the quantum entanglement concurrence $C$ is obtained
for even dimension pure states, with the upper and lower bounds of
$B$ governed by $C$.

\end{abstract}

 \maketitle

\section{Introduction}
Quantum nonlocality, such as that revealed by the violation of Bell
inequalities by quantum entangled states \cite{jsbe}, is one of the
most startling predictions of quantum mechanics. Recently, as
confirmed in loophole-free experiments \cite{bmlk}, nonlocality has
been proven to be useful in many quantum tasks such as
device-independent cryptography \cite{aanb} and randomness
certification \cite{spaa,rcol}.

According to Gisin theorem, all entangled pure states can display
nonlocal correlations \cite{ngis,syoc}, but some mixed entangled
states can provably satisfy local hidden variable models
\cite{rfwe,jbar,gtaa,mlsp}. Namely, there exist entangled mixed
states that never lead to nonlocality by any local POVM measurements
\cite{jbar}. Quantum nonlocality is usually associated with
entangled states that violate at least one of the Bell inequalities.
However, separable bipartite states can also show some nonlocal
properties \cite{chdp,shmb}. Despite all these progresses, the
precise relationship between entanglement and Bell violations has
remained less known, particularly for high dimensional bipartite and
multipartite cases.

Different ways to quantify the quantum nonlocality have been
presented \cite{jdcb,slsf,tzhy,eafp,abmn,sgba,afar,acsb,ecgg}, for
examples, the volume of the violation of Bell-type inequalities
\cite{eafp,abmn}. By employing the probability of violation of local
realism under random measurements \cite{ycnh}, in \cite{afar} the
authors investigated the nonlocality of entangled qudits with
dimensions ranging from $d=2$ to $d=10$. In \cite{acsb} the authors
proposed a machine learning approach for detection and
quantification of nonlocality. Quantifying Bell nonlocality by the
trace distance has been studied in \cite{sgba}.

For two-qubit states, the well-known CHSH inequality \cite{chsh} has
been used to detect the nonlocality. The corresponding operator is
given by $\mathbb{B}=A_1\otimes B_1+A_1\otimes B_2+A_2\otimes
B_1-A_2\otimes B_2$. The mean value $B\equiv\langle
\mathbb{B}_{\rho} \rangle$. For separable bipartite pure states
satisfies the CHSH inequality, $ B \leq 2$, where
$A_i=\vec{a}_i\cdot\vec{\sigma}$, $B_j=\vec{b}_j\cdot \vec{\sigma}$,
$\vec{a}_i$ and $\vec{b}_j$ are three-dimensional real unit vectors,
$i,j=1,2$, $\vec{\sigma}=(\sigma_1,\sigma_2,\sigma_3)$ with
$\sigma_1$, $\sigma_2$ and $\sigma_3$ the standard Pauli matrices.

In the following we use concurrence  $C$ \cite{wkoo,cbdd} as the
measure of quantum entanglement. Let $H_i$ denote the Hilbert space
associated with the $i$th subsystem. For a pure state
$|\psi\rangle\in H_1\otimes H_2$, the concurrence is defined by
\cite{Rungta01,AlbeverioFei01,kcsa}, $C(|\psi\rangle) =
\sqrt{2(1-Tr\rho_1^2)}$, where the reduced density matrix
$\rho_1=Tr_2|\psi\rangle\langle\psi|$ is obtained by tracing over
the second subsystem. The concurrence is then extended to mixed
states $\rho$ by convex roof,
$$
C(\rho)\equiv
\min\limits_{p_i,|\psi_i\rangle}\sum_ip_iC(|\psi_i\rangle),
$$
where the minimization goes over all possible ensemble realizations
$\rho=\sum_ip_i|\psi_i\rangle\langle\psi_i|$, $p_i\geq 0$ and
$\sum_ip_i=1$. For two-qubit states $C$ can be calculated directly
\cite{wkoo}. For high dimensional quantum states one has no general
results \cite{cjys,mjxn}.

For two-qubit states, the relationship between the concurrence $C$
and the mean value $B$ satisfy the relation: $2\sqrt{2}C\leq
B\leq2\sqrt{1+C^2}$ \cite{fvmw}. More recently, \cite{zsht} obtain
the necessary and sufficient condition that the upper bound can be
reached. Therefore, the Bell inequality is violated if
$C>1/\sqrt{2}$. Such relations have been also investigated by using
randomly generated two-qubit states \cite{xfzd}. The Bell inequality
for three-qubit states has been also studied \cite{jkan}. The
relationship between tripartite entanglement and genuine tripartite
nonlocality for three qubit Greenberger-Horne-Zeilinger class is
also investigated \cite{sgns}. The authors in \cite{pycs,pycc}
studied the relation between the upper bound of Bell violation and a
generalized concurrence for some n-qubit states. In \cite{qinhh} the
nonlocality distributions among multiqubit systems have been studied
based on the maximal violations of the CHSH inequality of reduced
pairwise qubit systems. Furthermore, from the reduced three-qubit
density matrices of the four-qubit generalized
Greenberger-Horne-Zeilinger (GHZ) states and W-states, a trade-off
relation among the mean values of the Svetlichny operators
associated with these reduced states has been presented
\cite{zhaolj}.

For high dimensional quantum systems less is known about the
relationship between concurrence and Bell violations. The main
difficulty lies in finding the mean value of suitable Bell
operators. In this paper, we explore the quantitative relationship
between concurrence $C$ and the Bell value $B$ for high dimensional
quantum systems.

\section{Bell nonlocality and concurrence of bipartite quantum states }
Based on Bell's idea \cite{jsbe}, for any given $n \times n$ real
matrix $N$ with entries $N_{ij}$, one can define a classical quantity,
\begin{equation*}
J(N)=\sup |\sum_{i,j=1}^{n}N_{ij}a_ib_j|,
\end{equation*}
where the supremum is taken over all possible assignment $a_i, b_j \in \{-1,1\}$,
$1 \leq i,j \leq n$. For any bipartite state $\rho$, the corresponding Bell
operator is defined by
\begin{equation*}
\mathbb{B}(N)=\sum_{i,j=1}^{n}N_{ij} A_i\otimes B_j,
\end{equation*}
where $A_i$ and $B_j$ are arbitrary observables whose absolute values of all eigenvalues are
less or equal to one.

A state $\rho$ is said to be nonlocal if it violates the following Bell inequality,
\begin{equation*}
B(N) \leq J(N),
\end{equation*}
where $B(N)= tr(\mathbb{B}(N)\rho)$ is the mean value of the Bell operator.
If one takes $N=\left(\begin{array}{cc}
 1 & 1 \\
 1 & -1
\end{array}\right)$, one gets the CHSH inequality with $J(N)=2$.

A pure $m\otimes n (m\leq n)$ quantum state has the standard Schmidt form,
\begin{eqnarray}\label{aaa}
|\psi\rangle=\sum_{i=1}^mc_i|a_ib_i\rangle,
\end{eqnarray}
where $c_i~(i=1,\cdots,m)$ are the Schmidt coefficients, and they
are in descending order, $|a_i\rangle$ and $|b_i\rangle$ are the
orthonormal bases in $H_1$ and $H_2$, respectively. The concurrence
of $|\psi\rangle$ is given by
\begin{eqnarray}\label{aab}
C=2\sqrt{\sum_{i<j}c_i^2c_j^2},
\end{eqnarray}
which varies from $0$ for pure product states to
$\sqrt{2(m-1)/m}$ for maximally entangled pure states \cite{kcsa}.

By selecting the matrix $N=\left(\begin{array}{cc}
 1 & 1 \\
 1 & -1
\end{array}\right)$
as that from the CHSH inequality, and Bell operators such as
$$A_0=I_{\frac{s}{2}}\otimes(\cos\theta\sigma_3+\sin\theta\sigma_1),$$
$$A_1=I_{\frac{s}{2}}\otimes(\cos\theta\sigma_3-\sin\theta\sigma_1),$$
$$B_0=I_{\frac{t}{2}}\otimes\sigma_3,$$
$$B_1=I_{\frac{t}{2}}\otimes\sigma_1,$$ for even $m,n$ or $$A_0=\left(\begin{array}{cc}
    I_{[\frac{s}{2}]}\otimes(\cos\theta\sigma_3+\sin\theta\sigma_1) &  0 \\
    0 & 1
  \end{array}\right),$$
$$A_1=\left(\begin{array}{cc}
    I_{[\frac{s}{2}]}\otimes(\cos\theta\sigma_3-\sin\theta\sigma_1) &  0 \\
    0 & 1
  \end{array}\right),$$
$$B_0=\left(\begin{array}{cc}
    I_{[\frac{t}{2}]}\otimes\sigma_3 &  0 \\
    0 & 1
  \end{array}\right),$$
$$B_1=\left(\begin{array}{cc}
    I_{[\frac{t}{2}]}\otimes\sigma_1 &  0 \\
    0 & 1
  \end{array}\right).
$$for odd $m,n$, then for
any bipartite pure state $|\psi\rangle$ as given by (\ref{aaa}), it
has been shown that \cite{ngap},
\begin{eqnarray}\label{bba}
B=2\sqrt{(1-\gamma)^2+K^2}+2\gamma,
\end{eqnarray}
where $K=2(c_1c_2+c_3c_4+\cdots)$, $\gamma=c_m^2$ for odd $m$, and
$\gamma=0$ for even $m$.

In the following we use (\ref{bba}) to obtain the following facts,
which does not depend on the optimality of (\ref{bba}).

\begin{theorem}
For any pure $m\otimes n (m\leq n)$ quantum state $|\psi\rangle$ as
given by (\ref{aaa}), we have
\begin{eqnarray}\label{bbb}
B\leq 2\sqrt{1+C^{2}}
\end{eqnarray}
for even $m$.
\end{theorem}

Proof: From \eqref{bba} we have for even $m$,
\begin{eqnarray*}
B=2\sqrt{1+K^{2}}.
\end{eqnarray*}
To prove the theorem , we only need to prove that
\begin{eqnarray}
C^{2}\geq K^{2},
\end{eqnarray}
namely
\begin{eqnarray}\label{ccc}
\sum_{i<j}c_{i}^{2}c_{j}^{2}\geq(c_{1}c_{2}+c_{3}c_{4}+\cdots)^{2}.
\end{eqnarray}

We prove (\ref{ccc}) by induction. For $m=2$, we have
\begin{eqnarray*}
C^{2}=4c_{1}^{2}c_{2}^{2}\geq K^{2}=4c_{1}^{2}c_{2}^{2},
\end{eqnarray*}
and the inequality \eqref{ccc} holds in this case.
Assume that the inequality \eqref{ccc} is true for $m=2(k-1)$ for arbitrary
positive integer $k$. For $m=2k$ we have
\begin{eqnarray}\label{ddd}
\sum_{i<j}^{2k}c_{i}^{2}c_{j}^{2}=\sum_{i<j}^{2(k-1)}c_{i}^{2}c_{j}^{2}
+\sum_{i}^{2k-2}c_{i}^{2}c_{2k-1}^{2}+\sum_{i}^{2k-1}c_{i}^{2}c_{2k}^{2}
\end{eqnarray}
and
\begin{eqnarray}\label{eee}
& &(c_{1}c_{2}+\cdots+c_{2(k-1)-1}c_{2(k-1)}+c_{2k-1}c_{2k})^{2}\nonumber\\
&=&(c_{1}c_{2}+\cdots+c_{2(k-1)-1}c_{2(k-1)})^{2}\nonumber\\
& &+2(c_{1}c_{2}+\cdots+c_{2(k-1)-1}c_{2(k-1)})c_{2k-1}c_{2k}\nonumber\\
& &+c_{2k-1}^{2}c_{2k}^{2}.
\end{eqnarray}
According to the assumption for $m=2(k-1)$, we have
\begin{eqnarray}\label{fff}
\sum_{i<j}^{2(k-1)}c_{i}^{2}c_{j}^{2}\geq(c_{1}c_{2}+\cdots+c_{2(k-1)-1}c_{2(k-1)})^{2}.
\end{eqnarray}
Therefore, we have
\begin{eqnarray}\label{ggg}
& &\sum_{i}^{2k-2}c_{i}^{2}c_{2k-1}^{2}+\sum_{i}^{2k-2}c_{i}^{2}c_{2k}^{2}\nonumber\\
&\geq&\sum_{j=1}^{k-1}c_{2j-1}^{2}c_{2k-1}^{2}+\sum_{j=1}^{k-1}c_{2j}^{2}c_{2k}^{2}\nonumber\\
&\geq&2(c_{1}c_{2}+\cdots+c_{2(k-1)-1}c_{2(k-1)})c_{2k-1}c_{2k},
\end{eqnarray}
where the second inequality is due to the inequality
$a^{2}+b^{2}\geq2ab$ for $a\geq0$ and $b\geq0$. According to the
inequalities \eqref{ddd}, \eqref{eee}, \eqref{fff} and \eqref{ggg}, we
obtain the inequality \eqref{ccc}, which completes the proof of the inequality \eqref{bbb}.

The upper bound of $B$ provided in inequality \eqref{bbb} consists with
the result of two-qubit case \cite{fvmw}. Therefore,
the inequality \eqref{bbb} can be regarded as the extension of the
relationship between $B$ and $C$ from two-qubit to high dimensional
quantum states. Besides, the inequality (\ref{bbb}) is saturated when only two $c_i$ are not zero.

Nevertheless, the lower bound of $B$ is generally
different from the one for two-qubit case. For high dimensional
systems we have the following general result.

\begin{theorem}
For any pure $m\otimes n (m\leq n)$ quantum state $|\psi\rangle$,
with the standard Schmidt form (\ref{aaa}), then for even $m$, we
have
\begin{eqnarray}\label{hhz}
B\geq\sqrt{2[1+C^{2}]}.
\end{eqnarray}
\end{theorem}

Proof: According to \eqref{aab} and \eqref{bba}, the inequality
\eqref{hhz} is equivalent to $1+2K^{2}\geq C^{2}$, namely, we need
to prove
\begin{eqnarray*}
1+2[4(c_1c_2+c_3c_4+\cdots+c_{m-1}c_m)^2]\geq
4\sum_{i<j}c_i^2c_j^2.
\end{eqnarray*}
That is, for odd $k,l$, we have
\begin{eqnarray}\label{iiii}
1+2[4\sum_{ k=1,l=1}^{m-1}(c_kc_{k+1}c_lc_{l+1})]\geq
2\sum_{i\neq j}c_i^2c_j^2.
\end{eqnarray}

Without loss of generality,
we assume that the Schmidt coefficients in (\ref{aaa}) satisfy $c_{i}\geq
c_{i+1},i=1,2,\cdots,m.$ Then we have the following facts,
$c_kc_{k+1}c_lc_{l+1}\geq c_{k+1}^2c_{l+1}^2$,
$c_kc_{k+1}c_lc_{l+1}\geq c_{k+1}^2c_{l+2}^2$,
$c_kc_{k+1}c_lc_{l+1}\geq c_{k+2}^2c_{l+1}^2$ and
$c_kc_{k+1}c_lc_{l+1}\geq c_{k+2}^2c_{l+2}^2$.
Hence, we have
\begin{eqnarray*} & &
4\sum_{k=1,l=1,odd}^{m-1}(c_kc_{k+1}c_lc_{l+1})\\
&\geq&\sum_{
k=1,l=1,odd}^{m-1}[(c_{k+1}^2c_{l+1}^2)+(c_{k+1}^2c_{l+2}^2)
\\& &+(c_{k+2}^2c_{l+1}^2)+(c_{k+2}^2c_{l+2}^2)]
\end{eqnarray*}
and
$$
1=(c_1^2+\sum_{i\neq 1}^mc_i^2)^2\geq 4c_1^2(\sum_{i\neq 1}^m c_i^2).
$$

Combining above relations we obtain
\begin{eqnarray*} & &
1+2[4\sum_{ k=1,l=1}^{m-1}(c_kc_{k+1}c_lc_{l+1})]\\
&\geq& 2\sum_{
k=1,l=1,odd}^{m-1}[(c_{k+1}^2c_{l+1}^2)+(c_{k+1}^2c_{l+2}^2)
\\& &+(c_{k+2}^2c_{l+1}^2)+(c_{k+2}^2c_{l+2}^2)]+4\sum_{i\neq 1}^m
c_1^2c_i^2\\& \geq & 2\sum_{i\neq j}c_i^2c_j^2,
\end{eqnarray*}
which gives rise to the inequality \eqref{iiii}, and proves
the inequality \eqref{hhz}.

From Theorem 2, it is obvious that if $C>1$, the state
$|\psi\rangle$ shows nonlocality.

\section{Conclusions and discussions}
Quantum nonlocality is a fundamental feature in quantum mechanics.
We have investigated the relation between the maximal expectation
value of Bell operators $B$ and entanglement concurrence $C$. The
upper and lower bounds of $B$ have been derived based on $C$. Such
the relations between $C$ and $B$ play important roles in judging
nonlocality from entanglement. Moreover, determining the
non-locality of high dimensional quantum states has been a difficult
problem in the theory of quantum information. Our results may
highlight further researches on the quantum nonlocality and related
the quantum correlations such as quantum steerability.

\bigskip
Acknowledgments:  This work is supported by the Natural Science
Foundation of Hainan Province(118QN230), the National Natural
Science Foundation of China under Grant Nos. 11861031, 11675113,
12075159, Beijing Municipal Commission of Education under grant No.
KZ201810028042, Beijing Natural Science Foundation (Z190005),
Academy for Multidisciplinary Studies, Capital Normal University and
Shenzhen Institute for Quantum Science and Engineering, Southern
University of Science and Technology (Grant No. SIQSE202005). This
project is also supported by the Academician Innovation Platform of
Hainan province and the Key Laboratory of data science and
intelligence education of Ministry of Education.

\bigskip
Declarations: All partial information is available.

\bigskip
Data Availability: Data sharing is not applicable to this article as
no new data were created or analyzed in this study.

\end{document}